# SpamDam: Towards Privacy-Preserving and Adversary-Resistant SMS Spam Detection

Yekai Li[1], Rufan Zhang[1], Wenxin Rong[1], and Xianghang Mi[1]

[1]*University of Science and Technology of China*
https://chasesecurity.github.io/SpamDam

*Abstract*—In this study, we introduce *SpamDam*, a SMS spam detection framework designed to overcome key challenges in detecting and understanding SMS spam, such as the lack of public SMS spam datasets, increasing privacy concerns of collecting SMS data, and the need for adversary-resistant detection models. *SpamDam* comprises four innovative modules: an SMS spam radar that identifies spam messages from online social networks(OSNs); an SMS spam inspector for statistical analysis; SMS spam detectors(SSDs) that enable both central training and federated learning; and an SSD analyzer that evaluates model resistance against adversaries in realistic scenarios.

Leveraging *SpamDam*, we have compiled over 76K SMS spam messages from Twitter and Weibo between 2018 and 2023, forming the largest dataset of its kind. This dataset has enabled new insights into recent spam campaigns and the training of high-performing binary and multi-label classifiers for spam detection. Furthermore, effectiveness of federated learning has been well demonstrated to enable privacy-preserving SMS spam detection. Additionally, we have rigorously tested the adversarial robustness of SMS spam detection models, introducing the novel *reverse backdoor attack*, which has shown effectiveness and stealthiness in practical tests.

## 1. Introduction

On the wheel of cellular networks and mobile devices, the short messaging service (SMS) made its debut in the 1990s for user-to-user communication , and has since then gained continuing popularity across the last few decades, which should be partly attributed to the adoption of SMS in critical applications such as business-to-customer communication (e.g., account alerts) and multi-factor authentication. However, the wide adoption of SMS is accompanied by SMS spam messages, i.e., unsolicited or even malicious messages that are distributed through SMS. Compared to other spamming channels such as Email, spamming through SMS has a much higher open rate of more than 90% while it is only around 20% for the Email channel [1]. Also, as SMS is commonly used for communication with critical business entities (e.g., bank services), it allows the spammers to masquerade as these legitimate parties and conduct various frauds and scams.

Particularly, SMS spam messages for phishing (i.e., smishing) in North America were reported to have increased 328% in Q3 2020 when compared to Q2 2020 [2], while more than 9 billion SMS spam messages had been blocked during 2022 by a security vendor for cellular users in China [3].

To address the issue of SMS spam, previous studies [4]–[10] have proposed various machine learning based detection systems, most of which abstract the detection of SMS spam as a binary text classification task along with various classification algorithms explored. However, these studies suffer from a set of limitations. First of all, *the groundtruth datasets for training and evaluation are outdated and small-scaled*. Particularly, the UCI spam dataset [4] was used in almost all detection studies, which, however, was collected in 2012 and contains only hundreds of SMS spam messages. As security prediction tasks are known to be vulnerable to the issue of concept drift [11], [12], it is unclear whether models trained on outdated datasets can achieve a satisfying performance when applied to real-world and up-to-date SMS spam campaigns. Also, *previous studies failed to explore fine-grained multiclass SMS spam classification*, which impedes not only an in-depth understanding of heterogeneous spam messages but also flexible enforcement of SMS spam countermeasures. Specifically, SMS spam messages can be further grouped into a set of categories, e.g., promotional spam, malware distribution, phishing/fraud spam, etc. And spam messages of different categories can vary significantly in the levels of security severity and should be further analyzed or alerted in a different manner.

Besides, as the enforcement of various regulations for privacy protection, it is becoming increasingly challenging or even prohibited to collect and store privacy-sensitive data from individuals, which is also applicable to SMS messages regardless of spam or non-spam. Then, another missing piece of previous studies is *how to collect SMS spam datasets in a privacy-preserving manner, and how to build up a SMS spam detection model in a privacy-preserving manner*.

Several studies have focused on assembling SMS spam datasets, each bearing distinctions from our approach. Reaves et al. [13] amassed approximately 400K messages through public SMS gateways, yet following a rigorous verification process, identified only 8.2% (8,178 messages) as actual SMS spam texts. In contrast, Gupta et al. [14] ventured into social media territory, gathering a vast collection of 70 million tweets to scrutinize phishing activities. Their analysis, however, was confined to phone numbers adopted in Twitter-based phishing, diverging from the scope of SMS spam as addressed in our study . Timoko et al. [15] introduced an innovative method by establishing a website that encourages users to voluntarily report spam messages. Despite the novelty of this live collection approach, the resultant dataset was notably limited, accruing just 85 spam messages over a span of two months. Besides, Tang, et al. [16] observed that SMS spam victims tend to complain about the received SMS spam messages on Twitter, and thus explored and demonstrated the potential of collecting such spam-reporting tweets from Twitter as well as extracting SMS spam messages from the spam-reporting tweets. However, as demonstrated in §3.1, the proposed *SpamHunter* degrades its performance significantly when applying to online social networks (OSNs) other than Twitter.

Lastly, despite applying machine learning to SMS spam detection, *no previous studies have comprehensively evaluated the adversary resistance of machine learning based SMS spam detection models*, which undermines the practicality of the proposed detection methods.

To address these limitations, we present in this study, *SpamDam*, an end-to-end framework to facilitate privacy-preserving and adversary-resistant SMS spam detection, which is made possible through a set of four novel modules as summarized below. The *SpamDam* is bootstraped by a module named *the SMS spam radar (SpamRadar)* which enables a continual discovery of SMS spam messages as reported on different OSNs. Then, given spam messages identified by the *SpamRadar*, the second module named *the SMS spam inspector* will analyze these spam messages along with their metadata attributes (e.g., the reporting time, and the natural language), in an attempt to gain a deep understanding of up-to-date SMS spam messages with regards to their scale, categories, and the temporal evolution, etc. Furthermore, multiple SMS spam classifiers will be built up and comprehensively evaluated through the third module named *the SMS spam detectors (SSDs)*. This module allows us not only to train multiple variants of a binary SMS spam classifier but also multi-class and multi-label SMS spam classifiers. Also, it enables us to evaluate, for the first time, the feasibility of federated learning for privacy-preserving training of SMS spam detection models. Then, given a set of SMS spam detection models, the fourth module named *the SSD analyzer* is designed to systematically evaluate the adversarial resistance of these models with regards to adversarial examples and poisoning attacks.

Leveraging *SpamDam*, we have discovered the largest-ever SMS spam dataset, built up a diverse set of SMS spam classification modules, as well as surfacing out a set of novel findings regarding both SMS spam messages and SMS spam detection. Below, we highlight some of these findings. In total, we have discovered 76,577 distinct SMS spam messages from two popular OSNs (Twitter and Weibo), which span the last five years between January 2018 and September 2023, are of tens of different natural languages, belong to diverse spam categories, and thus constitute the largest-ever and up-to-date SMS spam dataset.

Besides, upon this largest SMS spam dataset along with publicly available ones, we have built up a set of multi-lingual SMS spam classifiers. Particularly, our BERT-based binary spam classifier has achieved a state-of-the-art performance of 99.53% in recall and 99.28% in precision. On the other hand, we explore, for the first time, multi-label SMS spam classification and the resulting classifier has achieved a label ranking average precision (LRAP) score of 0.9281. Furthermore, we have also explored, for the first time, the feasibility of federated learning for privacy-preserving SMS spam detection. Experiments with realistic settings show that FL-trained SMS spam detection models can achieve a comparable performance when compared with centrally trained counterparts, while getting rid of the necessity of uploading any spam/non-spam datasets to the server.

Also, our model robustness evaluation reveals that models trained on outdated datasets are subject to significant concept drift, i.e., model aging. Particularly, a transformer-based model trained on the outdated UCI spam dataset has an initial recall of 99.03%, which decayed to less than 80% when evaluated on spam messages observed between 2018 and 2023. Furthermore, we demonstrate, for the first time, SMS spam detection models are vulnerable to testing-time realistic adversarial examples and training-time practical poisoning attacks. Particularly, attacks of imperceptible adversarial examples can achieve a success rate of 6.80% with only one perturbation and 20.00% with up to 5 perturbations. However, we also observe that the attack impact varies significantly across natural languages and spam categories, and adversarial training can help defend against such realistic adversarial examples. When it comes to training-phase poisoning attacks, under a practical poisoning rate ($\leq 5\%$), the untargeted poisoning attacks have minor attack impact on our SMS spam detection models, i.e., up to 1.60% drop in recall and almost no impact on the precision and false positive rate. We have also proposed and evaluated a novel variant of the backdoor attack wherein the attacker aims to deviate the spam detection model to mis-classify a benign message as spam as long as this message matching a pattern injected through spam reporting. And we name this attack as *the reverse backdoor attack* since the attack goal is not to evade spam detection but to demote benign messages, e.g., SMS messages distributed by a renowned organization. Our experiments show that such a reverse backdoor attack is both effective and stealthy. Particularly, it can achieve an attack success rate as high as 54.12% by poisoning as small as 1% SMS spam messages while the impact on the overall accuracy is just 1.58%, which strongly highlights the importance of data sanitization, especially for security detection systems (e.g., SMS spam detection) that depend on crowdsourced threat datasets.

Our contributions can be summarized as three-fold. First of all, we design and implement *SpamDam*, an end-to-end framework to facilitate privacy-preserving and adversary-resistant SMS spam detection. Besides, we have distilled a set of novel findings with regards to the landscape of up-to-date SMS spam, the effectiveness of multi-label SMS spam detection, the feasibility of applying federated learning to the training of SMS spam detection models, as well as the adversarial resistance of SMS spam detection, which can not only benefit future efforts on SMS spam detection but also inspire research on other security detection tasks. Lastly, Our study has led to the largest-ever SMS spam dataset and a set of SMS spam detection models, all of which will be released to the public in the short future.

## 2. Background and Related Works

**SMS spam detection.** The detection of SMS spam is typically defined as a binary classification task which takes the text content of a SMS message and optionally its attributes as the input and makes the decision regarding whether the SMS message is spam or not. To conquer this task, various machine learning algorithms have been explored, which range from traditional classification algorithms (e.g., naive Bayes, random forest, and support vector machine ) [4]–[6], to deep neural network architectures such as CNN [7], [8], LSTM and BiLSTM [7]–[9], and transformer models [10]. Also, various feature engineering techniques have also been proposed so as to extract features from the SMS text, such as tf-idf [6], word2vec [7], and various manual feature designs [4]. Furthermore, in addition to content-based features, non-content features have also been proposed and evaluated, such as temporal and topological features [5]. Particularly, Roy et al. [7] built up a CNN-based SMS spam detection model which has achieved a state-of-the-art performance when trained and evaluated on the UCI SMS spam dataset [4], and this model architecture is also evaluated in this study upon the largest-ever SMS spam dataset. Also, almost all these works focus on the binary classification of English SMS spam. Stepping forward, in this study, we evaluate, for the first time, not only *multilingual* SMS spam classification but also *multi-label* SMS spam classification.

One more limitation of these previous works is that almost all their detection models [4], [6], [7], [9], [10] were built upon the UCI SMS spam dataset [4] which, released in 2012, is small-scaled with only 747 English spam messages and more importantly is likely outdated. To address the scarcity of public SMS spam datasets, Tang et al. [16] proposed *SpamHunter* to collect SMS spam messages as reported by victims on Twitter, which results in the release of a multilingual dataset comprising 22K SMS spam messages, while Abayomi-Alli et al. [9] released another English spam dataset coined as *ExAIS_SMS* which was collected manually from 20 end users. However, as detailed in §3, the *SpamHunter* is heavily tailored for the Twitter platform, and fails to generalize to other social networks. These limitations motivate us to propose and build up *the SMS spam radar*, a self-contained spam collection system that enables cross-OSN SMS spam discovery with no dependency on any third-party commercial services.

**Black-box adversarial examples against text classification.** Compared with text classification tasks in other domains, security ones (e.g., SMS spam classification) tend to be deployed in a more adversarial environment and are more likely to be targeted by real-world adversarial attacks. In this study, for the first time, we will profile the resistance of SMS spam detection against realistic adversarial examples. Adversarial examples [17]–[19] are inputs crafted from authentic inputs through applying subtle perturbations with the intention to mislead a machine learning model and make its output deviate significantly from the correct one for the authentic input.

Among attacks of adversarial examples, some can be considered as white-box [20]–[22] as the attacker is assumed to have full access to the machine learning model including its architecture and parameters, whereas others belong to black-box attacks [21], [23]–[31] wherein only the outputs of a machine learning model is available for the attacker. Since black-box adversarial examples are more realistic compared with white-box ones, when evaluating the adversarial resistance of SMS spam detection models, we focus on black-box adversarial examples.

Existing attacks of black-box adversarial examples can differ notably in multiple aspects including the perturbation units, perturbation operations, and perturbation strategies. For instance, typical perturbation units include the character [23], the word [21], [26], [27], and the sentence [24], while exemplary perturbation operations are insertion [21], replacement [23], [25], [26], deletion [21], and paraphrasing [24]. Besides, various perturbation strategies have also been proposed, e.g., greedy algorithms [28], genetic algorithms [25], and various scoring functions [21], [23], [26]. Despite achieving a high attacking success rate, most attacks introduce perceptible changes to the original text, with one except being [31]. Boucher et al. [31] proposed the generation of adversarial examples through either inserting non-printable Unicode characters or replacing existing characters with homoglyphs so as to make the adversarial examples imperceptible. These non-printable Unicode characters include invisible characters, reordering characters, and deletion characters. When evaluating adversarial examples against SMS spam detection, we adopted these imperceptible adversarial attacks.

**Data poisoning attacks against text classification.** Different from adversarial examples which attack the testing phase of a model, data poisoning attacks [32]–[36] target the training phase and are referred as the injection of manipulated samples into the training data with the goal of undermining the resulting model's prediction performance for either all samples or samples of targeted classes. When the attacking goal is to degrade the overall performance, it is called the *untargeted* poisoning attack [37], [38]. Instead, a *targeted* poisoning attack [35] aims to make the resulting model mispredict samples of targeted classes but behave as normal for samples of other classes. Among targeted poisoning attacks, a well-known variant is the backdoor poisoning attack [36], [39] whereby the attacker pollutes the training data with backdoored samples so as to deviate the resulting model's prediction for samples matching the backdoor pattern, e.g., a unique visual element in an image, or a special sequence of characters in a text. Various poisoning attacks [40]–[43] have been proposed and evaluated for text classification models. However, very few studies have shed light to security prediction tasks which are inherently subject to more adversarial attacks including poisoning attacks. Stepping forward, we evaluate in this study the adversarial resistance of SMS spam detection models against multiple realistic poisoning scenarios.

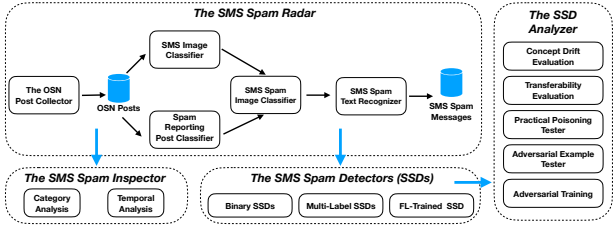

Figure 1. *SpamDam*: a framework to enable privacy-preserving and adversary-resistant SMS spam detection.

**Federated learning**. As an emerging distributed learning paradigm, federated learning (FL) [44] enables distributed clients (i.e., mobile devices) to jointly train a global deep learning model without the necessity of sharing their local privacy-sensitive data to any parties. In FL, a central server is deployed to instruct client-side local training, aggregate client-side model updates, and evaluate the resulting global model. A typical FL training process consists of rounds of client-side local training and server-side aggregation.

Also, FL is considered as *cross-device* when the clients are resource-constrained and large-scaled, e.g., tens of thousands of mobile devices. On the other hand, FL can be applied to the collaboration among large corporations (e.g., spam detection vendors), in which case, the clients are the participating organizations and thus of a smaller scale (e.g., $\leq 100$). Such a FL scenario is named as *cross-silo* FL. Although our SMS spam radar can help collect SMS spam messages from public OSNs, it is still privacy-invasive and thus impractical to collect benign SMS messages from end-user devices, not to mention uploading them to central server for model training. Therefore, in this study, we explore, for the first time, the feasibility of FL in terms of enabling privacy-preserving training of SMS spam detection models.

## 3. The Discovery and Characterization of SMS Spam

**An overview of the *SpamDam* framework.** As illustrated in Figure 1, *SpamDam*, our spam detection framework, consists of four modules. The first one is the SMS spam radar (*SpamRadar*) that is designed to continuously discover SMS spam messages as reported by victims on various OSN platforms. Given collected SMS spam messages, the SMS spam inspector is responsible to conduct statistical analysis and gain an in-depth understanding of SMS spam messages as well as the underlying spamming campaigns. The last two modules are the SMS spam detectors (SSDs) which enable both central training and FL-based training for a variety of SMS spam classifiers, and the SSD analyzer which evaluates the adversarial resistance of SMS spam classifiers. In this section, we first introduce how the *SpamRadar* is designed, and how the largest-ever SMS spam dataset was collected, and what statistical understandings we can distill from the collected SMS spam messages.

### 3.1. Privacy-Preserving SMS Spam Discovery

The only existing methodology that allows privacy-preserving and continuous discovery of SMS spam messages is *SpamHunter* [16]. However, it turns out to still suffer from two limitations, namely, constrained generalizability and dependency on third-party commercial services, which motivate us to build up our SMS spam radar (*SpamRadar*).

**Limitations of existing works**. Tang, et al. [16] observed that Twitter users may post tweets on the spam messages they have received, and proposed *SpamHunter* to automatically collect SMS spam messages as reported by Twitter users. However, it turns out that the *SpamHunter* fails to generalize well to other online social networks. Specifically, to evaluate the generalization of *SpamHunter*, we applied it to posts collected from Weibo, a Chinese online social network similar to Twitter. The manually crafted Weibo post dataset consists of 321 spam-reporting posts, 286 non-spam-reporting posts, 1,063 SMS screenshots as attached to Weibo posts, and another 887 images that are not SMS screenshots. The performance stats of *SpamHunter* for this Weibo dataset are listed in Table 1. And we can see that *SpamHunter* has a degraded performance on Weibo and it achieved an end-to-end recall of only 70.45% and a precision of only 52.54%, while they are 87% and 95% respectively on Twitter [16]. Similarly, the two underlying modules of *SpamHunter*, namely, the SMS image detector (SID) and the spam-reporting tweet classifier (SRTC), suffer from similar performance degradation. We further investigated the root causes of such performance degradation and the detailed explanations can be found in Appendix A.1. Particularly, when classifying whether a post is spam reporting or not, *SpamHunter* will first translate the post text into English, and we found that such a translation has led to the loss of subtle but critical semantics for many Chinese posts.

Furthermore, *SpamHunter* is not self-contained as it depends on two commercial 3rd-party services, for which more details can be found in Appendix A.1. In our study, we have made efforts to address these limitations, and the enhanced SMS spam collector is named as the SMS spam radar (*SpamRadar*).

**Overview of the SMS spam radar (*SpamRadar*).** Building upon the observation that victims of SMS spam can post received SMS spam messages on OSN platforms, our *SpamRadar* is designed to query multiple OSNs, identify spam reporting posts, and retrieve the embedded SMS spam messages if any. As illustrated in Figure 1, *SpamRadar* consists of five sub-modules. The first one is an OSN post collector designed to conduct keyword-based searches across OSN platforms and collect OSN posts relevant to SMS spam. Upon the OSN posts, an SMS image classifier (SIC) is built up to automatically decide whether a given image attached to an OSN post is a SMS screenshot or not. Furthermore, an extra multilingual text classifier is designed to profile the intention of an OSN post, i.e., whether an OSN post is intended for spam reporting, which is named as the spam-reporting post classifier (SRPC). Lastly, a separate classifier under the name of SMS spam screenshot classifier is designed to take the raw prediction results (probability values) of SSC and SRPC as the input and classify whether a given image attachment is a SMS spam screenshot. Once an OSN image is considered as an SMS spam screenshot, an extraction module will be applied to retrieve SMS spam messages from the screenshot.

TABLE 1. The performance of *SpamHunter* [16] on Weibo.

| Module | OSN | Accuracy | Precision | Recall |
|---|---|---|---|---|
| SID[1] | Tweet [16] | 95% | 98% | 93% |
| | Weibo | 88.81% | 89.55% | 72.50% |
| SRTC[1] | Twitter [16] | 88% | 95% | 90% |
| | Weibo | 52.72% | 51.48% | 97.55% |
| End-to-End | Twitter [16] | 91% | 95% | 87% |
| | Weibo | 84.57% | 52.54% | 70.45% |

[1] SID denotes the SMS image detector while SRTC denotes the spam-reporting tweet classifier.

TABLE 2. The groundtruth datasets for classification tasks in the SMS spam radar.

| Task | Data Group | Positive | Negative |
|---|---|---|---|
| Spam Reporting Post Classification | Twitter Original [16] | 500 | 250 |
| | Twitter New | 458 | 265 |
| | Weibo | 321 | 285 |
| | All | 1279 | 800 |
| SMS Image Classification | Twitter Original [16] | 687 | 313 |
| | Twitter New | 481 | 450 |
| | Weibo | 176 | 887 |
| | All | 1344 | 1650 |
| Spam Image Classification | Twitter | 199 | 201 |
| | Weibo | 77 | 163 |
| | All | 276 | 364 |

**The OSN post collector**. This collector is designed to gather spam-relevant posts from online social networks (OSNs). In a nutshell, the collector queries OSNs with keywords relevant to SMS spam. Each resulting post is made up of attributes that are commonly available across OSNs, e.g., the post text, the attached images if any, the post timestamp, the posting OSN account, among others. Also, the list of English spam keywords carefully crafted in [16] is reused here, e.g., *spam SMS*, and *smish SMS*. All these English keywords match the following pattern: {*malicious OR spam OR phish OR phishing OR smish OR scam OR fraud*} *SMS*.

Furthermore, unlike *SpamHunter* which queries Twitter with only these English keywords, our collector may translate the English spam keywords into multiple natural languages before querying an OSN, depending on what languages are popular on an OSN. For instance, Weibo is queried with the Chinese translations of these spam keywords, since the majority of Weibo posts are in Chinese and searching Weibo with English spam keywords typically returns almost nothing. Instead, when querying the Twitter platform wherein content can be distributed across tens of languages, the spam keywords are translated into 8 languages that are either of a large population of native speakers or very popular in spam distribution: Arabic, Chinese, English, French, Russian, Spanish, Indonesian, German. This multilingual searching strategy turns out to be effective in terms of improving the coverage of our searching. Particularly, when applying our collector to Twitter between January 2018 and September 2023, non-English keywords have exclusively triggered 30% resulting spam-reporting tweets and 27% spam messages. Also unlike the *SpamHunter* which supports only the Twitter platform, we have designed the collector in a way that is agnostic to the underlying OSNs, and it can be easily extended to new OSNs by adding OSN-specific crawling drivers. In the current implementation, it has support for both Twitter and Weibo.

**The spam-reporting post classifier (SRPC)**. Different from the original *SpamHunter* wherein a post text will be first translated into English before being classified by the spam-reporting tweet classifier, our SRPC is designed to directly take multilingual texts as the input, thus getting rid of any 3rd-party translation services. This is achieved through fine-tuning a multilingual pre-trained language model, namely, *the multilingual BERT* [1], which is pre-trained upon the Wikipedia corpus of 104 natural languages [45]. To train and evaluate SRPC, a groundtruth was composed from multiple sources. We first requested the original groundtruth dataset from the authors of *SpamHunter*, which consists of 500 spam-reporting tweets and 250 non-spam-reporting ones. We then manually labelled 723 extra posts from Twitter and 606 extra posts from Weibo. As a result, our final groundtruth contains 1,279 spam-reporting posts and 800 posts not intended for spam reporting. These posts are written in 27 different languages with 1,473 collected from Twitter and 606 collected from Weibo.

Given the groundtruth dataset, 80% were used during training while the remaining were used for testing. In addition to our multilingual model, we have also evaluated other machine learning options. One is a CNN model as adopted in [16] when building up the *SpamHunter*, and the other is an English BERT model [45]. For both, our multilingual groundtruth dataset is first translated into English before being used for training and testing. We also reproduced the spam-reporting Twitter classifier and evaluated it against our newly crafted testing dataset. As listed in Table 3, our multilingual BERT model has achieved the best performance in terms of recall (91.22%) while its precision (80.91%) is just slightly inferior to that of the BERT model. Since the

---

1. https://huggingface.co/bert-base-multilingual-cased

TABLE 3. THE PERFORMANCE COMPARISON AMONG DIFFERENT MODELS FOR SPAM-REPORTING POST CLASSIFICATION.

| Model | All-Test[1] | | Weibo-Test[2] | |
| --- | --- | --- | --- | --- |
| | Precision | Recall | Precision | Recall |
| SRTC[3] | 49.58% | 61.22% | 44.53% | 60.01% |
| CNN[4] | 63.33% | 88.76% | 56.45% | 85.52% |
| BERT[4] | 81.71% | 87.13% | 69.11% | 86.33% |
| Multilingual-BERT[4] | 80.91% | 91.22% | 68.37% | 95.21% |

[1] All-Test denotes the testing dataset combining posts from both Weibo and Twitter, constituting 20% of the groundtruth dataset.
[2] Weibo-Test refers to the Weibo posts in the testing dataset.
[3] SRTC denotes the spam-reporting tweet classifier proposed in [16].
[4] These models are trained on 80% of the groundtruth dataset.

TABLE 4. THE END-TO-END PERFORMANCE OF CLASSIFYING WHETHER AN OSN IMAGE IS A SMS SPAM SCREENSHOT.

| Testing Dataset | Model | Accuracy | Precision | Recall |
| --- | --- | --- | --- | --- |
| Twitter | SIC & SRPC[1] | 87.4% | 95.2% | 74.4% |
| | Random Forest | 95.6% | 94.9% | 94.9% |
| Weibo | SIC & SRPC | 89.8% | 85.2% | 83.6% |
| | Random Forest | 92.8% | 88.6% | 89.7% |
| All | SIC & SRPC | 88.6% | 90.3% | 78.4% |
| | Random Forest | 94.2% | 92.2% | 92.6% |

[1] denotes the *AND* strategy wherein an image is considered as a SMS spam screenshot only when the predictions of both SRPC and SIC are positive.

SRPC is only used to generate features for the SMS spam image classifier, this performance turns out to be sufficient. Also, when increasing the probability threshold from 0.5 to 0.89, our multilingual model can achieve a higher precision of 96.64% at the overhead of a lower recall of 44.75%, which suggests our SRPC can also be used independently for filtering posts. Besides, we also looked into the false predictions of SRPC and more details are presented in Appendix A.2.

**The SMS image classifier (SIC).** When reporting a SMS spam, OSN users tend to include the screenshot of the received spam message, in an attempt to easily present the spam message in its original format. In *SpamRadar*, given an OSN post relevant to SMS spam, another module is built up to automatically decide whether an attached image is a SMS screenshot or not. Different from [16] which considered it as an objection detection task for identifying SMS cells, we abstract it as an image classification task and the resulting module is named as the SMS image classifier (SIC).

Considering transformer-based models have achieved SOTA performance in many image classification tasks [46], we adopted the paradigm of pre-training and fine-tuning wherein the ViT model [2], a transformer model pre-trained on ImageNet-21k was selected for fine tuning. To train and evaluate SIC, images attached to raw OSN posts were sampled and labeled, which leads to a groundtruth dataset of 1,344 SMS screenshot images and 1,650 non-SMS images as listed in Table 2. Given this groundtruth dataset, 80% were used to fine-tune the ViT model while the remaining were for evaluation. As a result, the model has achieved a recall of 96.77%, a precision of 94.74%, and a F1-score of 95.74%.

**The SMS spam image classifier (SSIC).** Given the classification results of SRPC and SIC, an image can be considered with high confidence as a spam screenshot as long as both models output positive predictions, i.e., the image is classified as a SMS screenshot and the respective OSN post is classified as spam-reporting. This simple *AND* strategy is what was adopted in [16]. However, our evaluation reveals that this strategy will miss a non-negligible portion of spam messages, which is mainly attributed to the observation that many OSN posts don't convey affirmative intention regarding spam reporting, even if the attached image is a SMS spam screenshot. OSN posts with such an ambiguous intention will likely be classified by SRPC as negative (not spam reporting), in which case the attached image will not be considered as spam screenshots.

To address this issue, a 2nd-level binary classifier is built up in replacement of the simple *AND* strategy. This classifier takes the probability outputs of both SRPC and SIC as the inputs, and outputs whether an OSN image is a spam screenshot or not. We name this classifier as the SMS spam image classifier (SSIC). When training SSIC, OSN posts and images used in training either SIC or SRPC were excluded so as to avoid any cyclic dependency issues. Otherwise, if an OSN post used to train SRPC was reused to train SSC, the post would be predicted by the SRPC that was trained on itself. Instead, we collected and labeled another groundtruth dataset that is disjoint to that for SRPC and SIC, and it consists of 276 SMS spam screenshots belonging to 252 OSN posts and 364 non-spam images belonging to 237 OSN posts. Our SSC was implemented as a random forest classifier with 15 trees and the maximum tree depth being 12. Its evaluation performance is listed in Table 4 along with a comparison with the simple *AND* strategy. And we can see, compared with the simple *AND* strategy, our SSC has increased the precision from 90.3% to 92.2%, and more importantly, the recall from 78.4% to almost 92.6%.

**The spam text recognizer (STR).** Going through the aforementioned modules, SMS spam screenshots along with their parental OSN posts will be identified, upon which, we move to extract the SMS spam texts. Different from *SpamHunter* which relies on a third-party commercial service to extract text from images, our spam text recognizer is built upon Tesseract [47] [3], an open source OCR engine that can recognize text of more than 100 languages from images of various formats. When applying Tesseract, we have designed multiple pre-processing and post-processing steps so as to achieve a higher recognition rate. As a result, our STR has achieved an overall recognition accuracy of 94.4%, a word accuracy of 97% and a character accuracy of 99%. For more details regarding STR especially the processing steps, please refer to Appendix A.3.

---

2. https://huggingface.co/google/vit-base-patch16-224

3. https://github.com/tesseract-ocr/tesseract

TABLE 5. THE PERFORMANCE OF *SpamRadar* ON DIFFERENT OSNS.

| Testing Dataset | Accuracy | Precision | Recall |
|---|---|---|---|
| Twittwer | 95.5% | 97.8% | 92.7% |
| Weibo | 99.5% | 93.3% | 100.0% |
| Reddit | 85.5% | 97.7% | 83.0% |
| Xiaohongshu | 88.5% | 84.0% | 92.3% |

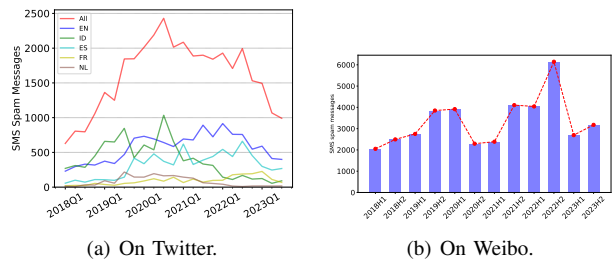

(a) On Twitter.   (b) On Weibo.

Figure 2. The temporal evolution of SMS spam messages as observed on both Twitter and Weibo. For 2023 H2, we only have data for the first three months (July to September).

**End-to-End performance.** To assess the real-world effectiveness, the while *SpamRadar* pipeline was applied to posts collected via the OSN-specific collector from both Weibo and Twitter. Given posts predicted through the pipeline, 200 posts per OSN were randomly sampled for manual validation. As shown in Table 5, our pipeline exhibits robust performance for both OSN platforms. On Twitter, it achieves an end-to-end precision of 97.8%, i.e., 97.8% images predicted as SMS spam screenshots are true cases. On the other hand, the precision slightly drops to 93.3% on Weibo.

**Generalization to other OSNs.** As previously discussed, *SpamRadar* is designed to be OSN-agnostic, i.e., it can be extended to new OSNs with OSN-specific crawling drivers. This has been further demonstrated through our collection and analysis of data from Reddit and Xiaohongshu, thereby affirming the generalization of our pipeline.

Given the constraints presented by the Reddit APIs, we resorted to manually identifying spam-related posts that contain images, using the same meticulously crafted keywords employed in collection of Twitter/Weibo posts. For Xiaohongshu, we developed a specialized crawling driver to gather similar spam-reporting posts, adhering to the same keyword set. Given these posts, the spam radar was applied, which was followed by manual validation for 200 sampled predictions for each OSN. As shown in Table 5, the *SpamRadar* pipeline has achieved a decent performance for both OSN platforms. Particularly, a high precision of 97.7% (127 out of 130 predicted SMS spam screenshots) is achieved for Reddit while the recall of 83.0% (127 out of 153 true SMS spam screenshots) is still acceptable. On one hand, these results demonstrate the generalization of the *SpamRadar* to different OSNs. On the other hand, fine tuning with OSN-specific ground truth can still further enhance the pipeline, e.g., improving the precision for Xiaohongshu.

**Deployment.** We then deployed *SpamRadar* to collecting SMS spam messages from both Weibo and Twitter in November 2023 and set up the searching time period between January 2018 and September 2023. As the results, we have discovered a total of 76,577 distinct SMS spam messages, among which 39,909 (52.12%) were sourced from 36,366 Weibo posts and 33,870 Weibo accounts while the left 36,668 were extracted from 32,747 tweets and 27,071 Twitter accounts. To the best of our knowledge, this is the largest-ever SMS spam dataset that has been collected and will be released to the public. Below, we present fine-grained measurements on this dataset, which is achieved through another module of our *SpamDam* framework, namely, the SMS spam inspector.

### 3.2. Understanding Up-To-Date SMS Spam

**The SMS spam inspector.** Given up-to-date SMS spam messages discovered from OSNs, it is valuable to understand their landscape in terms of scale, distribution, and evolution. The SMS spam inspector is designed to fulfill this goal. In a nutshell, it takes SMS spam messages and spam-reporting OSN posts as the input, and outputs an extendable set of statistical data points surrounding the given SMS spam dataset. Below, we illustrate how it works and what statistical results it can generate through the measurements on the newly discovered SMS spam dataset.

**SMS spam reported on Twitter**. In total, 36,668 SMS spam messages and 32,747 spam-reporting tweets have been discovered by *SpamRadar* for the last five years between January 2018 and September 2023. Compared to *SpamHunter* [16], our *SpamRadar* have discovered 67% more SMS spam messages. Even considering only the same period (January 2018 and December 2021), 18% more spam messages have been collected by our *SpamRadar*. On the other hand, 6,017 spam messages ever captured by *SpamHunter* were missed by our pipeline, which is mainly due to the non-existence of the respective tweets.

*The temporal distribution.* We also profiled the temporal distribution of SMS spam messages, for which, the timing of a SMS spam message is approximated using the timestamp of the respective OSN post, under the assumption that the time gap between the report of a SMS spam message and the recipient of the message is small. Figure 2(a) presents such a temporal evolution for SMS spam messages discovered on Twitter. We can observe that SMS spam messages have been continuously reported on Twitter at a notable volume. Specifically, the volume of SMS spam trends upward between 2018 Q1 and 2020 Q2, resides on the plateau for two years, before starting to slowly go downward in 2022 Q2, which is consistent with previous observations in [16] as well as our extra observation for SMS spam messages discovered on Weibo. Also, the high volume of SMS spam messages reported between 2020 Q1 and 2022 Q2 is likely related to the Covid-19 pandemic, as also observed in some other studies [48], [49].

*The language distribution.* In total, SMS spam messages reported on Twitter are distributed in 54 natural languages,

TABLE 6. THE CATEGORIES OF SPAM MESSAGES REPORTED ON WEIBO.

| Category | Subcategory | Spam Ratio | Labelled Messages |
|---|---|---|---|
| Promotion | General Promotion | 40.25% | 353 |
| | Gambling | 2.74% | 24 |
| | Sex&Porn | 3.53% | 31 |
| Fraud | Finance | 36.49% | 320 |
| | Account Alarm | 4.33% | 38 |
| | Insurance | 2.74% | 24 |
| | Delivery | 6.27% | 55 |
| | Acquaintance | 1.03% | 9 |
| | Prize | 2.62% | 23 |

with the top ones as English (35%), Indonesian (26%), Spanish (20%), French (6%), Dutch (5%). Compared to [16], a lower fraction (35%) of English spam messages has been observed, which is 42% observed in [16]. On the other hand, the ratio of Spanish spam has increased by 10% when compared with that observed in [16]. Besides, Chinese spam messages account for a very small fraction (0.03%) of spam messages collected on Twitter. On contrary, 95.94% spam messages captured from Weibo are in Chinese, which further highlights the necessity of cross-OSN SMS spam discovery.

*Categories.* To understand the subcategories of spam messages as collected from Twitter, 500 SMS spam messages were sampled out for manual labeling. Similar to the observations of [16], most spam messages are intended to fraud or phish the victims and are thus masqueraded as various activities such as online account alerts, tax refund, shipping updates, etc. To facilitate understanding of spam categories, we also explored how to automatically classify a spam message into pre-defined spam categories, and the details will be presented in §4.2.

**SMS spam reported on Weibo.** Through applying the *SpamRadar* to the Weibo platform, 39,909 distinct SMS spam messages have been captured, which were attached in 36,366 spam-reporting posts across the period between January 2018 and September 2023.

*The temporal distribution.* As shown in Figure 2(b), SMS spamming activities in China are trending upwards with a concerning volume. Particularly, the number of SMS spam messages reported on Weibo has increased from 2,051 in 2018 H1 to 6,144 in 2022 H2. Also, over 3K spam messages have already been observed for the first three months of 2023 H2, which is even higher than that of the whole 2023 H1. Besides, one thing to note, the trend experienced a trough in 2020 H2 and the following 2021 H1, but bounced back to the peak in 2022 H2, which may be related to the number of COVID-19 infections in China, which was at a low level at 2020 H2 and 2021 H1, but surged again since then until 2022 H2 [50]. Also, during that period, the more COVID-19 infections there were, the more restrictions on traveling and outdoor activities, which likely contributed to the surge of spamming activities and spam-related reports on Weibo.

*Language distribution.* As discussed above, spam messages reported on Twitter are widely distributed across many different languages. On contrary, only 10 languages were observed for spam messages reported on Weibo, and Chinese dominates the distribution with a large fraction of 95.94%, which aligns with the fact that Weibo is known to be mainly used in China.

*Categories.* Despite being mostly in Chinese, spam messages reported on Weibo are diverse in terms of their categories. To learn their categories, we randomly sampled and manually labeled 877 SMS spam messages based upon their content. As shown in Table 6, almost 53% spam messages are intended to *harm* the victims through fraud and phishing, while the left 47% are unsolicited promotion messages. Such a category-wise distribution aligns with that of Twitter-reported spam messages. Also, many Weibo-reported spam messages are composed with text semantics that are closely related to Chinese culture or local services in China. For more details, please refer to Appendix B.

## 4. The Classification of SMS Spam

Given the largest-ever SMS spam dataset collected, SMS spam detection is well explored from multiple aspects, as illustrated in Figure 1. Particularly, we have explored not only the well-studied binary SMS spam classification along with novel model architectures, but also fine-grained multi-label SMS spam classification. Furthermore, the feasibility of training a SMS detection model through federated learning (FL) has been demonstrated for the first time, which allows the training of a spam detection model without uploading any SMS spam/non-spam messages from clients to the server and is thus more privacy-preserving.

### 4.1. Binary SMS Spam Classification

Most previous works on SMS spam classification are dedicated to English messages. However, one spam message can contain characters of multiple languages, and language translation can incur non-negligible performance overhead (§3.1). We thus proceed to explore the feasibility of multilingual SMS spam classification which takes raw multilingual SMS spam messages as input and outputs the binary classification results. To achieve this goal, we have adopted the emerging machine learning paradigm of pre-training and fine-tuning, which has achieved state-of-the-art performance in many NLP classification tasks [45], [51], [52]. As a result, a set of binary classification models have been built up, which vary in their training and testing datasets, as detailed later. Across these models, spam detection is considered as a downstream classification task upon a pre-trained large language model (LLM), and the specific LLM chosen in our study is the multilingual BERT language model [4] which is trained on top 104 languages with largest Wikipedia corpus. Next, we will introduce these models with a focus on describing their groundtruth datasets and comparing their spam detection performance.

---

4. https://huggingface.co/bert-base-multilingual-cased

TABLE 7. THE GROUNDTRUTH FOR SMS SPAM CLASSIFICATION.

| Dataset | Period | Spam | Non-Spam | All |
|---|---|---|---|---|
| UCI | 2012 | 747 | 4,827 | 5,574 |
| ExAIS | 2015 | 2,350 | 2,890 | 5,240 |
| Twitter | 2018-2023 | 36,668 | 0 | 36,668 |
| Weibo | 2018-2023 | 39,909 | 0 | 39,909 |

**Groundtruth datasets.** Table 7 lists the four SMS spam datasets to be used to train and evaluate our spam classification models. Two of them are collected by our *SpamRadar* from Weibo and Twitter respectively, while the other two, namely UCI [5] and ExAIS [6], are publicly available SMS spam datasets and are commonly used in previous works for evaluating SMS spam detection [4]–[10]. Among these datasets, the two collected by *SpamRadar* contribute most spam messages but none non-spam messages, while the left two complement the groundtruth dataset with not only all the non-spam messages but also spam messages collected before 2016.

For each of the four datasets, 80% are randomly sampled out to compose the training dataset (e.g., $Train_{UCI}$ and $Train_{Twitter}$), while the left 20% are used for testing (e.g., $Test_{Weibo}$ and $Test_{ExAIS}$). Given these datasets, we move to train four binary SMS spam classifiers, as listed below.

$BERT_{All}$. This is the classification model trained on training segments of all the aforementioned four groundtruth datasets, i.e., $Train_{All}$. Here, we trained this spam detection through fine-tuning the aforementioned BERT multilingual language model. Also, during the training, we set the number of epochs as 20, the batch size as 32, and the max text length as 80.

$CNN_{All}$. Roy et al. [7] proposed and implemented a CNN model for spam detection along with SOTA performance achieved on the UCI dataset. We reproduced this CNN model but on the training dataset $Train_{All}$, the same training dataset used in $BERT_{All}$. One thing to note, the original CNN architecture doesn't support multiple languages, instead, samples will be first translated into English when necessary, before being fed into the model for training and evaluation.

$BERT_{All-Weibo}$ and $BERT_{All-Twitter}$. Sharing with $BERT_{All}$ the same model architecture and training parameters, $BERT_{All-Weibo}$ differs in that its training data excludes samples from Weibo. Similarly, another model $BERT_{All-Twitter}$ is built up upon all training samples except ones from Twitter. These two models are designed to profile whether detection knowledge learned upon spam messages of one OSN can be well applied to the detection of spams reported on the other OSN.

Given the four models, we evaluated them on three different testing dataset. One is all the test data $Test_{All}$ which is composed by test segments of all aforementioned four groundtruth datasets. Then, to take a closer look into the effectiveness of $BERT_{All-Weibo}$ on spams collected from

5. https://archive.ics.uci.edu/ml/datasets/SMS+Spam+Collection
6. https://github.com/AbayomiAlli/SMS-Spam-Dataset

TABLE 8. A PERFORMANCE COMPARISON AMONG BINARY SMS SPAM CLASSIFICATION MODELS.

| Model | $Test_{All}$ | | |
|---|---|---|---|
| | Precision | Recall | False Positive Rate |
| $CNN_{All}$ | 98.86% | 98.83% | 10.05% |
| $BERT_{All}$ | **99.28%** | **99.53%** | **7.49%** |
| $BERT_{All-Twitter}$ | 98.68% | 99.29% | 13.75% |
| $BERT_{All-Weibo}$ | 99.24% | 99.45% | 7.92% |

TABLE 9. THE PERFORMANCE ON **IMBALANCED** TEST DATASET.

| Model | Accuracy | Precision | Recall | FPR [1] |
|---|---|---|---|---|
| $BERT_{All}$ | 98.89% | 99.28% | 99.53% | 7.49% |
| $BERT_{weighted}$ | 97.77% | 97.75% | 99.82% | 21.38% |
| $BERT_{balance}$ | 96.36% | 99.53% | 96.42% | 4.25% |

[1] FPR stands for the false positive rate.

Weibo, we exclude Twitter samples from $Test_{All}$, which leaves spams reported on Weibo dominating the left test data, and we thus name it as $Test_{All-Twitter}$. Similarly, to evaluate the effectiveness of $BERT_{All-Twitter}$ on spam messages reported on Twitter, $Test_{All-Weibo}$ is composed.

A direct performance comparison among the four models are listed in Table 8. As we can see that BERT-based classification models have achieved better performance than the previously SOTA CNN model. Particularly, when evaluating on $Test_{All}$, our $BERT_{All}$ model has increased the recall by 0.7% and the precision by 0.42% while decreasing the false positive rate by 2.56%.

Another key observation is that SMS spams reported on Twitter contribute most to the performance of $BERT_{All}$. Particularly, when evaluated on $Test_{All}$, the model $BERT_{All-Twitter}$ trained without Twitter-reported SMS spam messages has the recall (99.29%) lower by 0.24%, the precision lower by 0.60%, and the false positive rate higher by 6.26%, when compared with $BERT_{All}$ (99.53%). Instead, the model $BERT_{All-Weibo}$ trained with Twitter-reported spams but not Weibo-reported spams has almost the same performance with $BERT_{All}$. A reasonable explanation is that SMS spam samples reported on Weibo are mostly distributed in China and cannot well represent the data distribution of globe-wide SMS spam campaigns. Thus, the model ($BERT_{All-Twitter}$) trained mostly on Weibo-reported spam messages cannot well capture Twitter-reported global spam messages. We also compared the performance of these models when evaluated on $Test_{All-Twitter}$ and $Test_{All-Weibo}$, and more details can be found in Appendix C.1.

TABLE 10. THE PERFORMANCE ON BALANCE TEST DATASET.

| Model | Accuracy | Precision | Recall | FPR[1] |
|---|---|---|---|---|
| $BERT_{imbalance}$ | 64.19% | 58.35% | 99.17% | 70.79% |
| $BERT_{weighted}$ | 57.71% | 54.18% | 99.82% | 84.41% |
| $BERT_{balance}$ | 97.84% | 99.24% | 96.42% | 0.74% |

[1] FPR stands for the false positive rate.

**The effect of ground-truth imbalance.** As the default groundtruth dataset is imbalanced with more spam messages (Table 7), we also profile its effect on model performance through comparative analysis. To achieve this, two extra BERT models are built up. One is BERT$_{weighted}$ which differs from BERT$_{All}$ in that a weighted loss function is employed during training. The other is BERT$_{balance}$ which is trained on a balanced groundtruth rather than the default one. To construct such a balanced dataset, the default ground truth is further augmented with wild tweets as non-spam that are sampled from the Twitter Archiving Project[7]. This is based upon the assumption that most regular tweets are non-spam. Such an assumption is not only consistent with insights gleaned from prior research [53], [54], but is also aligned with our manual evaluation of 1,000 randomly selected wild tweets, as 97% of them are confirmed to be benign.

Table 9 presents the performance of these BERT models when evaluated upon the test part of the default ground truth, i.e., the imbalanced test dataset. We also evaluate these models on the test part of the balanced groundtruth, as presented in Table 10. We can see that the BERT$_{balance}$ exhibits superior performance across both balanced and imbalanced test datasets. Notably, in addition to good performance with regards to recall and precision, it achieves low false positive rates of 4.25% and 0.74% for both test datasets, which are much lower than its two counterparts. Conversely, BERT$_{All}$ records a prohibitively high false positive rate of 70.79% when assessed against the balanced test dataset, a testament to its limited exposure to non-spam messages during training, thereby impairing its ability to accurately identify non-spam messages. Besides, BERT$_{weighted}$ also records even worse false positive rates, which suggests the weighted loss function didn't yield an improvement in performance. An actionable lesson we can take is that, to train a robust SMS spam classifier, it is a necessity to incorporate diverse non-spam messages. Also, when non-spam SMS messages are not sufficient, wild tweet texts can be a good substitution.

**A comparison with publicly available anti-spam options.** We also conduct a comparative analysis between our models with three anti-spam counterparts: OOPSpam [55], Perspective [56], and zero-shot in-context learning with GPT-4. The first two services are renowned for their content-based filtering capabilities, offer APIs that process text inputs to return detection outcomes. Also, OOPSpam categorizes texts as either "spam" or "ham", while Perspective assigns a probabilistic score ranging from 0 to 1, with a threshold of 0.5 employed to ascertain spam content. Besides, GPT-4, one of the most advanced generative large language models, is known to be a good few-shot or zero-shot learner [57]. To make GPT-4 a zero-shot in-context learner for SMS spam filtering, a prompt (Appendix C.2) is carefully designed by following well-adopted prompt engineering practices [58]–[60].

To facilitate this evaluation, 100 spam messages are randomly sampled from the test dataset along with 100

7. https://archive.org/details/twitterarchive

TABLE 11. THE PERFORMANCE OF DIFFERENT SPAM DETECTION CANDIDATES.

| Anti-spam | Accuracy | Precision | Recall | FPR[1] |
|---|---|---|---|---|
| GPT-4 | 89.5% | 89.1% | 90.0% | 11.0% |
| OOPSpam | 61.5% | 83.3% | 29.7% | 6.1% |
| Perspective | 83.0% | 75.4% | 98.0% | 32.0% |
| BERT$_{All}$ | 95.0% | 90.9% | 100.0% | 10.0% |
| BERT$_{balance}$ | 97.0% | 96.1% | 98.0% | 4.0% |

[1] FPR stands for the false positive rate.

non-spam messages. Considering some anti-spam candidates (e.g., Perspective, OOPSpam) don't support multilingual input, each test message is first translated into English before being evaluated against all these anti-spam candidates.

The comparative performance results are summarized in Table 11. Among all the three anti-spam candidates except for ones implemented by our study, GPT-4 demonstrates superior accuracy in spam detection, while OOPSpam exhibits a low false positive rate of 6.1%, albeit at the cost of a low recall, suggesting a propensity to classify SMS messages as non-spam. Conversely, Perspective reports the highest false positive rate, indicating a tendency towards generating false alarms, aligning with findings reported in [16]. Compared to public anti-spam options, the models developed in this study exhibit superior performance in both accuracy and precision. This enhancement is likely due to the specialized focus on SMS spam messages, which is not the primary target for many generic anti-spam services, resulting in their less satisfactory results. Specifically, when examining the BERT$_{All}$ and BERT$_{balance}$ models, the latter demonstrates the benefits of incorporating a balanced dataset. BERT$_{balance}$ not only achieves the highest precision of 96.1% but also maintains a significantly lower false positive rate of 4.0%. These results underscore the importance of including diverse-enough non-spam texts when training anti-spam models.

### 4.2. Multi-Label SMS Spam Classification

Previous studies on SMS spam detection focus on *binary* spam classification. However, spam messages can reside in different categories which can vary significantly in their security severity levels, i.e., different levels of maliciousness. For instance, a fraud spam message is more harmful than a promotion one, and the more harmful a spam message, the more severe alert it should trigger along with more attention from anti-spam practitioners. Also, in real-world spam fighting, it is valuable to learn the category distribution of spam messages as well as how spam activities of harmful categories evolve across time, for which manual analysis is becoming increasingly infeasible. To fill in this gap, we proceed to build up a multiclass SMS spam classifier and a multi-label SMS spam classifier, both of which have achieved promising performance and thus paved the lane for fine-grained detection and analysis of SMS spam.

**SMS spam categories.** Through manually looking into SMS spam cases, we have defined 10 spam categories that are either most commonly observed in our spam dataset or reported in previous studies [16]. As listed in Table 12, the category of *promotion* encompasses most unsolicited promotional or marketing spam messages except for gambling spam and spam promoting sex & porn, for both of which, we define separate categories under the name of *P-Gambling* and *P-Sex*. In addition to these three promotion categories, we also define 7 fine-grained categories for fraudulent spam messages, depending on their semantic topics. Specifically, Epidemic messages refer to pandemic-related fraud spam,particularly those related to COVID-19, whose category is defined as *F-COVID*. Besides, fraudulent messages masqueraded under the hood of financial topics will be assigned in the category of *F-Finance*, while *F-Account* specify fake messages intended to steal your online account credentials. Other fraudulent categories include *F-Insurance* for insurance scams, *F-Delivery* for forged shipping update messages, *F-Acquaintance* for scams pretending to be the victim's acquaintance , and *F-Prize* for prize or lottery scams. One thing to note, we are not intended to exhaust all spam categories or to identify the best definitions of spam categories. Instead, we focus on evaluating the feasibility of multi-class/multi-label SMS spam classification.

**Data labeling.** To minimize the subjective impact on the quality of data labeling, a systematic and robust data labeling process is devised, involving the collaboration of three annotators. This process comprises four key stages. Initially, we analyze the data slated for classification to establish the primary categories and define labeling criteria. Subsequently, each of the three individuals independently undertakes the complete data labeling task. Following this, any data with inconsistent labels undergoes analysis to refine category definitions and quantities, and to reaffirm labeling criteria. The three individuals then independently re-label this subset of data. Finally, for any remaining data with label inconsistencies, the three individuals engage in collective discussion to reach a consensus.

**A multi-class SMS spam classifier.** In our multi-class SMS spam classifier, each message is predicted as one of aforementioned 10 categories. To build up this classifier, we have labeled a groundtruth dataset of 2,017 SMS spam messages, among which, 1,140 were captured from Twitter and 877 were from Weibo. The category distribution of these labeled messages is shown in Table 12. As you can see, this dataset is not balanced, and some categories (e.g., Promotion and F-Finance) have hundreds of samples while other categories (e.g., P-Gambling, and F-COVID) have only tens of samples. To accommodate this, we use a weighted loss function when training our neural network model, and the loss for mis-classifying a sample $s$ of category $c$ is inversely proportional to the fraction of samples of category $c$ in the groundtruth dataset, i.e. the categories with fewer samples will be assigned more weights when calculating the loss, and vice versa.

Similar to the binary SMS spam classifier, we choose the paradigm of pre-training and fine-tuning, especially considering our groundtruth dataset is of a small size. Then, the same BERT multilingual language model is selected as the pre-trained model. During the fine-tuning, the number of epochs was set as 10, with the batch size as 16, the learning rate as 1.207e-5, and the maximum sequence length as 128. The resulting model was trained on 80% of our groundtruth dataset. Evaluation upon the left 20% samples revealed a micro recall of 87.95% and a micro precision of 87.95%. Category-specific performance metrics are listed in Table 12. As we can see, multi-class SMS spam classification is promising. Particularly, the category-wise precision ranges from 83.33% for F-Finance to 96.00% for F-Acquaintance.

TABLE 12. THE PERFORMANCE OF MULTI-CLASS SMS SPAM CLASSIFICATION.

| Spam Category | % Groundtruth | Precision | Recall | F1-Score |
|---|---|---|---|---|
| Promotion | 28.21% | 88.57% | 90.47% | 89.23% |
| P-Gambling | 1.49% | 88.69% | 88.58% | 88.63% |
| P-Sex | 1.64% | 86.66% | 86.11% | 86.30% |
| F-Finance | 28.16% | 88.33% | 76.67% | 79.81% |
| F-Account | 24.24% | 83.37% | 86.00% | 84.27% |
| F-Insurance | 1.59% | 94.40% | 92.46% | 93.32% |
| F-Delivery | 7.79% | 89.36% | 96.67% | 92.63% |
| F-Acquaintance | 1.04% | 96.00% | 73.33% | 80.97% |
| F-COVID | 1.39% | 95.00% | 81.00% | 86.11% |
| F-Prize | 4.96% | 87.68% | 89.37% | 88.48% |

When labeling the groundtruth, we found out that some spam messages may belong to multiple categories, which is echoed by the prediction results of the multi-class SMS spam classifier. For example, there is a spam: *"LLOYDS : You have successfully scheduled a payment of 54.99GBP to a new payee for 18:45 PM on 14/12 . If this was NOT you visit : https : // removepayees.net/lloyds"*. It was labeled as *F-Account* by our annotators, while the multi-class classifier model predicted it as F-Finance. Indeed, it is a fraud related to both account alarms and financial updates. Another spam *"QUICK LOANS only use vehicle letters (MOBILL only) rates start from 0.7 % without survey & BI C3K help takeover GERRY 081210015656 bit.ly/2DyzOVU"* was labeled as *F-Finance*, but the multi-class classifier model predicted it as *Promotion*, which is also reasonable to some extent. Therefore, we are motivated to explore the feasibility of classifying each spam message into multiple categories, i.e., multi-label spam classification.

**A multi-label SMS spam classifier.** To facilitate a direct comparison with the multi-class classifier, the same set of samples were used to compose the groundtruth. Then, for each sample, in addition to the single label assigned in multi-class classification, more labels would be assigned if it indeed belongs to multiple categories. As the results, among the 2017 spam samples, 618 have been assigned with two or more labels, while 33 have further three or more labels.

Following the multiclass classifier, the same model architecture as well as training parameters were utilized to train this multi-label classifier. The resulting model has achieved a label ranking average precision (LRAP) score of 0.9281 whereby the LRAP [61] is a well-acknowledged metric for multi-label classification along with the best score being 1. Another metric we consider is Hamming score. Considering $N$ samples in the testing dataset, assume $Y_i$ denotes the set of labels for the $i^{th}$ sample, and $\hat{Y}_i$ denotes the predicted labels for the same sample. Then, the Hamming score is calculated as $\frac{1}{N} \sum_{i=1}^{N} \frac{|Y_i \cap \hat{Y}_i|}{|Y_i \cup \hat{Y}_i|}$, and a Hamming score of 1 denotes a model has got 100% predictions right. Then, when evaluated on the testing dataset, our multi-label spam classifier has achieved a Hamming score of 0.7940. We also measured the category-wise precision and recall, as shown in Appendix C.3. Overall, we can conclude that both multi-class and multi-label SMS spam classification tasks are promising and feasible, despite a non-negligible variance across spam categories.

TABLE 13. THE PERFORMANCE OF FL-TRAINED BINARY SMS SPAM CLASSIFIERS.

| Training Type | Dirichlet | Precision | Recall | Accuracy |
|---|---|---|---|---|
| Central (BERT$_{All}$) | N/A | 99.28% | 99.53% | 98.91% |
| Cross-Device FL | $\alpha = 0.5$ | 99.09% | 99.45% | 98.73% |
| | $\alpha = 1$ | 99.01% | 99.35% | 98.56% |
| | $\alpha = 10$ | 98.88% | 99.20% | 98.29% |
| Cross-Silo FL | $\alpha = 0.5$ | 99.24% | 99.43% | 98.80% |
| | $\alpha = 1$ | 99.27% | 99.53% | 98.91% |
| | $\alpha = 10$ | 99.31% | 99.53% | 98.92% |

### 4.3. SMS Spam Detection via Federated Learning

Aforementioned classification tasks assume a central training scenario wherein all training/testing datasets are available on the central server. However, this central training scenario is becoming increasingly impractical for SMS spam detection, as SMS messages tend to be privacy-sensitive and uploading SMS messages (spam or not) to the central server can violate various privacy protection regulations. Next, We present our experiments of applying federated learning (FL), a privacy-preserving training paradigm, to SMS spam detection, in an attempt to address the tension between SMS spam detection and user data protection. More background knowledge of FL can be found in §2.

**The FL experiment settings.** Flower [62], a popular FL framework is adopted in our FL experiments. Also, our experiments evaluate both cross-device FL and cross-silo FL. For cross-device FL experiments, we assume the total number of clients $N$ is 200 while it is 20 for the cross-silo FL. Then, regarding what fraction of FL clients $n$ will be sampled in each FL round, we set it as 10% for cross-device FL and 100% for cross-silo FL. This is reasonable since participants of the cross-silo FL for SMS spam detection tend to be security vendors and should be always available for FL training, while participants of cross-device FL are likely to be mobile devices which have a very limited availability. In addition, the well-adopted FedAVG [44] is chosen as the default aggregation rule. Then, the client-side FL settings consist of a learning rate $l = 5e^{-5}$, a batch size $\beta = 32$, and $e = 2$ local epochs.

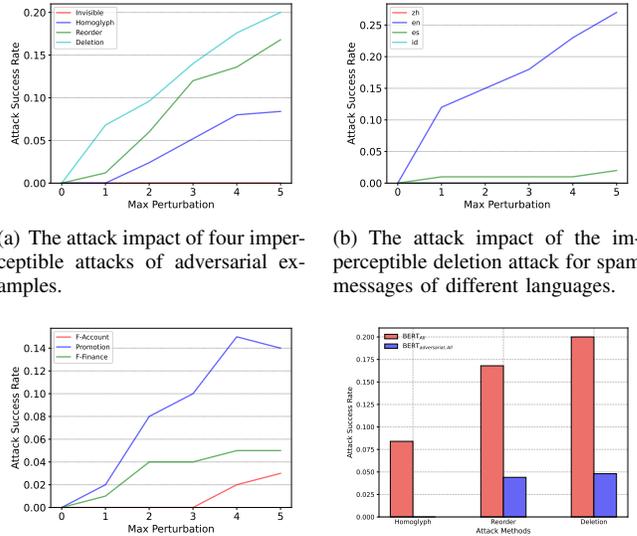

(a) The attack impact of four imperceptible attacks of adversarial examples.

(b) The attack impact of the imperceptible deletion attack for spam messages of different languages.

(c) The attack impact of the imperceptible deletion attack for spam messages of different categories.

(d) The defensive impact of adversarial training.

Figure 3. The effectiveness of adversarial examples and adversarial training on SMS spam detection.

Then, aforementioned binary SMS spam classifier BERT$_{All}$ is selected as the baseline model. Also, similar to central training, the same set of 20% groundtruth are held out for testing on the FL server, while the left 80% will be distributed to the $N$ clients by following a Dirichlet distribution [63] with the default alpha value of $\alpha = 1$, which gives a quantity-imbalanced distribution of the samples across the clients. In our experiments, instead of exhaustively exploring various FL hyperparameters such as the local batch size, we focus more on evaluating the impact of realistic non-IID data distributions of spam/non-spam samples. To simulate different extent of quantity-based non-IID data distribution, the Dirichlet distribution [63] with the following alpha values will be explored: $\alpha \in \{0.5, 1, 10\}$ wherein the larger the $\alpha$ is, the more balanced the distribution of samples across FL clients is.

**The FL experiment results.** Table 13 presents the performance stats for FL-trained binary SMS spam classifiers under different quantity-based Dirichlet distributions. And we can see that the performance of FL-trained models is comparable to that of the centrally trained counterparts. More details about the FL training process can be found in Appendix C.4.

## 5. The Adversarial Resistance of SMS Spam Classification

As revealed and evaluated in a long line of works [21], [23]–[31], text classification models can be vulnerable to a variety of adversarial attacks, particularly, training-stage poisoning attacks, and inference-stage adversarial examples. Being a text classification task in the security domain, SMS spam detection is more likely to be targeted by these adversarial attacks. In this section, an extensive set of experiments have been conducted to profile the adversarial robustness of SMS spam classification with a focus on training-stage poisoning attacks and testing-state adversarial examples.

## 5.1. Adversarial Examples

When evaluating the robustness of SMS spam classification models against adversarial examples, we adopt the text perturbation techniques proposed in Boucher et al. [31]. As detailed in §2, compared to previous studies, the key advantage of [31] resides in the imperceptibility of the resulting adversarial examples, i.e., the algorithms introduce imperceptible mutations to the original text, maintain the stealthiness while misleading the targeted text models, and are thus considered as more realistic. This is achieved through either inserting non-printable Unicode characters or replacing existing characters with homoglyphs, and these non-printable Unicode characters can be grouped into three categories, which include invisible characters, reordering characters, and deletion characters. Also, to decide which characters to insert under a given budget (the maximum number of perturbations), differential evolution [64], a gradient-free optimization method is deployed.

To evaluate our SMS spam detection (SSD) models against these imperceptible attacks, the original implementation [8] of these attacks from the authors has been integrated into our SSD analyzer. Also, the aforementioned transformer-based binary SMS spam classifier (BERT$_{All}$) is chosen as the target model, and 250 authentic spam samples of 14 different languages have been randomly selected out for testing. Figure 3(a) presents how the attacking success rate varies across the four perturbation strategies and different attacking budgets. And we can see that attacking via inserting deletion control characters turns out to be most effective, which can achieve an attacking success rate of 6.80% with only one injection and 14.00% with up to 3 injections. However, the injection of invisible characters has no attacking impact, regardless of the maximum number of injections, which aligns with the original work [31].

We also observed that the attack success rate can vary significantly when encountering spam messages of different languages or different categories. Figure 3(b) presents how the success rate of the imperceptible deletion attacks varies across four languages with most spam messages observed. And we can see that the imperceptible deletion attack is only effective for English/Spanish spam messages, but has zero success rates for both Chinese and Indonesian, which is applicable to all the three effective imperceptible attacks. Then, when it comes to different spam categories, as shown in Figure 3(c), these attacks work best for the spam category of Promotion, which is expected since promotional spam messages tend to reside closest to the decision plane between spam messages and non-spam ones. Particularly, under the attack budget of 5 perturbations, the imperceptible deletion attack has achieved a success rate of 15% for spam messages belonging to Promotion while it is only 3% for F-Account.

---

8. https://github.com/nickboucher/imperceptible

**Adversarial training.** Given some adversarial example attacks are effective, we move to explore the effectiveness of adversarial training in terms of hardening the SMS spam detection models. Specifically, we first sampled out 5% SMS spam messages from the overall groundtruth, which are exclusive to aforementioned 250 SMS spam messages for testing adversarial examples. Then, we applied the three effective imperceptible attacks (excluding the invisible attack due to its ineffectiveness) to these sampled spam messages with a maximum perturbations of 5, which generated 1,125 effective adversarial examples. We then utilized these adversarial examples along with the original training dataset for BERT$_{All}$ to train BERT$_{adversarial,All}$, which achieved a performance comparable to BERT$_{All}$, when evaluated on the original testing dataset.

We then evaluated the robustness of this adversarially trained spam detection model, using the aforementioned 250 SMS spam messages. As illustrated in Figure 3(d), adversarial training has significantly lowered the attacking success rates for all the three imperceptible attacks. For instance, adversarial training has lowered the success rate of deletion attack from 20.0% to 4.80%. More details can be found in Appendix D.1.

## 5.2. Poisoning Attacks

Different from adversarial examples, poisoning attacks focus on manipulating the training data rather than perturbing testing data. And the attack goal is to undermine the resulting model's prediction performance for either any test sample (untargeted poisoning) or samples of targeted classes (targeted poisoning) or patterns (backdoor attacks). As our spam detection pipeline involves discovering SMS spam messages from OSNs, it gives attackers the opportunity to poison the training dataset via posting well crafted spam reporting posts on OSNs, which motivates us to evaluate the robustness of SMS spam detection models under realistic poisoning attacks.

**The threat model.** We assume a typical poisoning attacker is capable of posting spam reporting posts across OSNs. Also, such posts can be published through many OSN accounts, e.g., a scale of hundreds. Also, we assume the attacker has no knowledge of who is collecting SMS spam reports from OSNs or any details of the SMS spam detection models. On the other hand, we assume it is unlikely for a poisoning attacker to pollute a non-negligible volume of benign SMS messages, as benign SMS messages are sourced from end users and it is costly to control a large portion of end users and manipulate their local SMS messages.

**Poisoning scenario I: the untargeted poisoning attack.** In this scenario, we assume the attacker can only poison spam samples through reporting benign messages as spam. Also, the fraction of spam messages that are maliciously reported by the attacker is defined as $p$ and our experiments vary with different $p$ values. In our experiments, 80% of the original groundtruth dataset (§4) are used for training along with poisoned ones, while the left 20% are free of poisoning and held out for performance evaluation.

TABLE 14. THE ATTACKING IMPACT FOR UNTARGETED POISONING ATTACKS UNDER DIFFERENT POISONING RATES $p$.

| Metrics | The Poisoning Rate | | | | | | |
|---|---|---|---|---|---|---|---|
| | 0 | 1% | 5% | 10% | 45% | 50% | 60% |
| Accuracy | 98.91% | 97.52% | 97.47% | 97.13% | 97.45% | 97.27% | 96.56% |
| Precision | 99.28% | 99.13% | 99.26% | 99.27% | 99.31% | 97.30% | 96.48% |
| Recall | 99.53% | 98.11% | 97.93% | 97.54% | 97.86% | 99.74% | 99.83% |
| FPR[1] | 7.49% | 7.99% | 6.84% | 6.70% | 6.34% | 25.77% | 33.91% |

[1] FPR is short for the false positive rate.

Then, there is a challenge regarding where to get benign messages before mis-reporting them as SMS spam. Here, upon the assumption that most posts on Twitter are benign messages, we randomly sample tweets from a large-scaled tweet dataset, namely, the Twitter Archiving Project [9], and consider these tweets as a good approximate for benign SMS messages. Given the training dataset and a poisoning rate of $p$, randomly sampled tweets will be injected into the training dataset as spam messages only if they are predicted as non-spam by the *authentic* spam detection model (BERT$_{All}$).

Table 14 illustrates how the poisoned spam detection model varies in its performance when evaluated under different poisoning rates, while the results of more poisoning rates can be found in Appendix D.2. As we can see, a *practical* untargeted poisoning attack can degrade the recall performance to a notable extent but has almost no impact on the precision. Specifically, even the injection of only 1% poisoned samples ($p = 1\%$) can degrade the recall by 1.42%, but the impact on the precision is within the margin of error. Then, as the $p$ is over 50% which is considered impractical, we start to see a notable drop in precision, and more importantly, the significant increase in the false positive rate. Particularly, as the $p$ has increased from 45% to 50%, the false positive rate has jumped from 6.34% to 25.77%. Regarding what $p$ is practical, as aforementioned in §3.1, the 76,577 messages in our SMS spam dataset were sourced from 60,941 distinct OSN accounts, and we consider a practical poisoning attacker that has control of up to hundreds of OSN accounts. We thus consider a practical poisoning rate as $p \leq 5\%$. Under such a practical range of the poisoning rate, we argue that this untargeted poisoning attack has minor attack impact on our SMS spam detection model, i.e., up to 1.60% drop in recall and almost no impact on the precision and false positive rate.

**Poisoning scenario II: reverse backdoor attack through reporting stamped benign messages as spam.** Then, we consider a variant of the backdoor attack wherein the attacker aims to deviate the spam detection model to mis-classify a benign message as spam as long as this message matches a pattern injected through spam reporting. In the real world, such a scenario can occur when the attacker tries to demote benign messages distributed by a legitimate

9. https://archive.org/details/twitterarchive

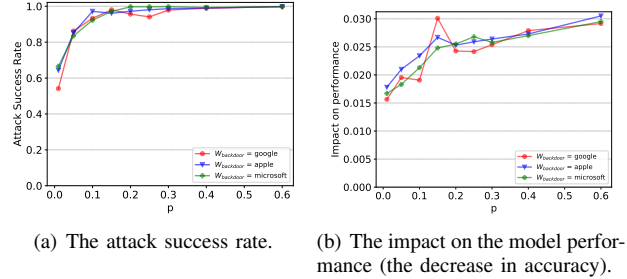

(a) The attack success rate.
(b) The impact on the model performance (the decrease in accuracy).

Figure 4. The effectiveness of reverse backdoor attacks via injecting stamped **benign** messages under the label of spam.

organization (e.g., a banking service). We name this attack as *the reverse backdoor attack*, since the attack goal is to deviate classification for benign messages of victim parties rather than spam messages of malicious parties.

To profile this attack, we first explored the reporting of backdoored benign messages as spam at different poisoning rates ($p$). Also, like poisoning scenario I, tweet text messages randomly sampled from the Twitter Archiving Project were first filtered by the authentic spam detection model, and the ones considered as non-spam would be used as the original benign messages to backdoor. Since we consider the attacking intention of demoting benign messages distributed from a specific brand, e.g., Google, a set of well-known brand names were chosen to compose the set of backdoor words $W_{\text{backdoor}}$. Then, one experiment was conducted for each distinct combination of $p$ and $w_{\text{backdoor}}$, e.g., the experiment for $p = 10\%$ and the backdoor word "google". Across these experiments, the randomly selected tweet messages will first be injected with a single backdoor word $w_{\text{backdoor}}$ and the injection position is decided in a random manner. Then, some of these stamped messages will be polluted into the training dataset under the spam label, while the left 1,000 will be used to test the backdoored model.

Figure 4(a) presents the backdoor effect in terms of mis-classifying benign testing messages that are stamped with the respective backdoor word (e.g., *google*), while Figure 4(b) presents the impact of these backdoor attacks on the overall model performance as gained from evaluation on the held-out testing dataset. And we can conclude that this reverse backdoor attack is very effective in terms of misleading the model to falsely alarm stamped but benign messages as spam, while maintaining a low impact on the overall model performance. Particularly, when the $p$ is just as low as 1%, the attack success rate can be as high as 54% while the impact on the overall accuracy is just 1.58%. Then, when the $p$ is larger than 15%, the poisoned model tends to mis-classify all stamped benign messages as spam.

**Poisoning scenario III: reverse backdoor attack through reporting stamped spam messages.** The third scenario differs with the 2nd one in that it injects true spam messages rather than benign messages, which gives rise to a even more stealthy attack since the injected messages are indeed spam and can thus likely bypass potential human vetting. Then, regarding experiment settings, this scenario shares the same sets of $p$ and $W_{\text{backdoor}}$ with the second one. However, the injection process is different. Specifically, given the poisoning rate $p$, the backdoor word $w_{\text{backdoor}}$, and the number of spam messages in the training dataset $N_{\text{spam}}$, $N \times p$ spam messages will be randomly sampled out from the training dataset and get stamped with $w_{\text{backdoor}}$ through random word injection. These stamped true spam messages will be put back and thus poison the training dataset. Then, 1,000 benign tweet messages that are randomly selected out from the Twitter Archiving Project will also be stamped with the same backdoor word and will then be used to test the attacking impact.

TABLE 15. THE IMPACT OF THE TWO CLASSES OF REVERSE BACKDOOR ATTACKS ON THE OVERALL MODEL PERFORMANCE IN TERMS OF THE DECREASE IN ACCURACY.

| Class | The Poisoning Rate | | | | | |
|---|---|---|---|---|---|---|
| | 1% | 5% | 10% | 15% | 20% | 40% |
| Benign | 0.0167 | 0.0196 | 0.0213 | 0.0272 | 0.0250 | 0.0274 |
| Spam | 0.0156 | 0.0157 | 0.0165 | 0.0175 | 0.0163 | 0.0200 |

Similar to poisoning scenario II, backdooring via reporting stamped spam messages can achieve a high success rate when the poisoning rate is practically low (e.g., $p \leq 5\%$). Along with the high backdoor success rate is the low impact on the overall performance, which is no more than 2% in accuracy decrease when $p \leq 30\%$. Figures profiling such a pattern can be found in Appendix D.3. Table 15 also presents a direct comparison between backdooring with benign messages and backdooring with spam messages, with regards to their impact on overall model performance. As we can see, backdooring with spam messages consistently incurs a lower impact on model performance, and is thus more stealthy than backdooring with benign messages. By now, we can conclude that SMS spam detection built upon crowdsourced SMS spam messages is vulnerable to practical reverse backdoor attacks.

## 6. Discussion

**The adversarial resistance of SMS spam detection.** As revealed in §5, practical adversarial text examples can achieve a notable success rate against SMS spam detection. Also, given a practical poisoning rate, the reverse backdoor attacks can effectively and stealthily compromise the model's performance for stamped but benign messages while maintaining a low fingerprint on the overall model performance. As the defense measures, adversarial training appears to be a necessity. Besides, to defend against the effective reverse backdoor attacks, a whitelist of brand names may be enforced to sanitize the collected spam messages before using them for training. Such observations also have strong implications for other security prediction systems, especially ones built upon crowdsourced threat datasets. Particularly, the crowdsourced data points should be well sanitized before being fed into the training pipeline.

**Ethical considerations and actions.** When discovering SMS spam messages from OSNs, we strictly follow the data crawling policies of the respective OSN platform. Also, the collected OSN posts and the extracted SMS spam messages are securely stored and we only distill statistical data points from these raw data rather than looking for any privacy-related information. We thus believe our study incurs no ethical concerns or risks.

**The concept drift and the model transferability.** Given the extensive observation of concept drift in other security prediction tasks [12], e.g., malware detection [11], it is interesting and valuable to evaluate the extent to which SMS spam detection is subject to this issue. Our results on concept drift can be found in Appendix E. We have also profiled the transferability between SMS spam detection and relevant tasks including Email spam classification and general toxic text classification. In another word, we want to figure out *whether a SMS spam classifier is applicable to Email spam classification and toxic text classification, and vice versa.* For more details, please refer to Appendix F.

**Data and code release.** To facilitate future anti-spam efforts, We plan to release the source code of *SpamDam*, along with the resulting SMS spam datasets and the SMS spam detection models[10]. One thing to note, the SMS spam messages will be carefully processed before released, so as to remove any personally identifiable information relevant to the respective spam-reporting OSN users.

## 7. Conclusion

As shown in this study, SMS spam messages can be collected at a large scale without incurring privacy concerns. Also, fine-grained multi-label SMS spam classification is promising and the training of SMS spam detection models can also be privacy-preserving via federated learning. However, on the other hand, SMS spam classification models are found to be vulnerable to adversarial examples, which calls for the necessity of adversarial training. Also, depending on crowdsourced SMS spam messages can introduce a novel poisoning surface (the reverse backdoor attacks), which calls for extra measures to be explored and adopted. We believe all these observations are valuable in terms of enabling privacy-preserving and adversary-resistant SMS spam detection as well as informing other security detection tasks especially ones that are either privacy-sensitive or have dependence on crowdsourced datasets.

---

10. https://github.com/ChaseSecurity/SpamDam

# Appendix A.
# More Details on SMS Spam Discovery

## A.1. Limitations of Previous Studies on Collecting SMS Spam from OSNs.

We further investigated why the *SpamHunter* performed not well when applied to Weibo, through looking into its false predictions. The first observation is that many false negative cases of the SID are due to one or more of the following factors: the dark background of the SMS image, smaller-than-usual SMS boxes, long SMS texts, etc. On the other hand, most false positives of the SID can be attributed to the fact that they have frames resembling SMS boxes. Furthermore, when classifying whether a post is spam reporting or not, the post text will first be translated into English before feeding into the SRTC. We found out that the translation leads to the loss of subtle but critical semantics for some Chinese posts and thus more misclassification cases in terms of both false positives and false negatives. Such misclassified Chinese posts include interrogative posts, rhetorical questions, implicit negative sentences, etc.

Besides, *SpamHunter* is not self-contained as it depends on two commercial 3rd-party services. One is the Google

Translation service, which is utilized to translate a non-English tweet text into English before being further classified. The other is an OCR API[11] from Google Cloud which is used to extract text out of an SMS spam screenshot. However, depending on third-party cloud services prevents *SpamHunter* from being deployed in an offline mode. Besides, it also prevents future works from accurate reproduction of the performance results as these commercial services are subject to changes in APIs and underlying mechanisms.

## A.2. False Predictions of the Spam-Reporting Post Classifier

We also looked into false negative cases of the spam-reporting post classifier when the threshold is 0.89. As shown in Figure 5(a), the original text in Chinese deliberately conveys the poster's anger for receiving spam messages. Although a human annotator can understand the intrinsic meaning of the post, the post may appear with a positive sentiment for a machine learning model, which misleads the classifier. The same also goes for the case in Figure 5(c). Besides, when it comes to a post with an interrogative tone (Figure 5(b)), the classifier can fail to decide whether it's a spam reporting post or not.

```
Original text:        现在的垃圾短信还怪贴心的
Translated to English: The spam messages are so sweet now
Probability:          [0.1671, 0.8329]
```

(a) False negative case 1.

```
Original text:   @lunomoney this is one of the phishing sms right?
Probability:     [0.3253, 0.6747]
```

(b) False negative case 2.

```
Original text:        哈哈哈哈哈哈哈果然是垃圾短信、突然看到的垃圾短信感觉还挺好玩的
                      hhhhhhhh感觉被骂的好爽
Translated to English: Hahahahahahahaha, it really is a spam text message,
                      the spam text message I saw suddenly felt quite amusing
                      hhhhhhhh I feel so good being scolded
Probability:          [0.2069, 0.7931]
```

(c) False negative case 3.

Figure 5. False negative cases of the SRPC.

## A.3. The SMS Spam Text Recognition

Based upon the observation that Tesseract doesn't work well for SMS screenshots with a dark background. Before applying Tesseract, an SMS spam screenshot will be first pre-processed to detect if it has a dark background. And if it has, an extra pre-processing step will be taken to change the dark background to a bright one. Given the screenshot, a heuristic-based method is designed to decide whether it has a dark background. Specifically, the screenshot will first be read in grayscale. If a pixel has a pixel value less than 30, it is considered as a dark pixel. Finally, we compute the ratio of dark pixels to all pixels in the screenshot, and the screenshot will be identified as a dark image if the ratio is larger than 70%. Once a screenshot with a dark background is identified, we inverse the color of the image, which means

11. https://cloud.google.com/vision/docs/ocr

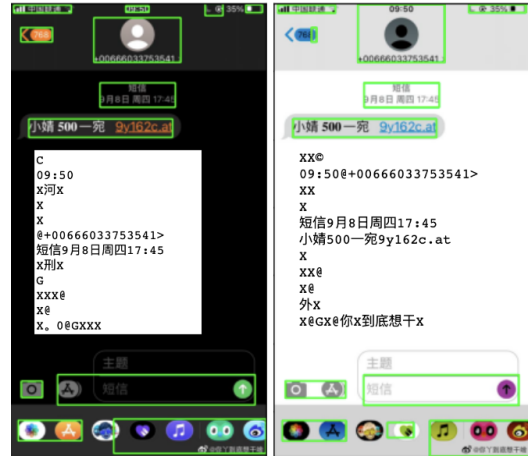

Figure 6. A comparison between a dark-background image and the pre-processed counterpart along with the OCR result.

we will change the original pixel value $x$ to $255-x$. Figure 6 presents a dark-background SMS spam screenshot, and the pre-processed counterpart, along with the respective OCR results.

Also, after applying Tesseract, a data dictionary will be returned for each SMS screenshot, which contains the recognized text cells, coordinates and confidence for each recognized text cell. Then, a set of post-processing steps turn out to be necessary to merge together text cells belonging to the same SMS message as well as distinguishing the spam message from other text elements, e.g., the sender's phone number, and the receiving datetime. In a nutshell, these post-processing steps involve multiple threshold-based heuristics. Specifically, given word cells identified by Tesseract, a threshold-based merging method is applied to merge word cells belonging to the same SMS spam message. This threshold is 1.5 times the current word box, which means two word cells will be merged if the distance between them is less than such a number. Also, some SMS spam messages have been mosaicked by the spam reporter, likely to hide privacy-sensitive text elements (the victim's name), which leads to wrong characters in the recognized SMS spam message. To address this issue, a simple but effective strategy is applied, which replaces a recognized letter or word with a white space when the respective recognition confidence score is lower than 65. Also, multiple regular expression are applied to filter out text elements that are not the SMS spam message, such as the sender's phone number, the datetime string, etc.

To verify the performance of STR, we randomly sampled 1,000 SMS images analyzed by STR, with 500 from Twitter and 500 from Weibo, and manually identified the correct SMS messages from their screenshots. Among these 1,000 images predicted by the SSIC as SMS spam screenshots, 52 are false positives (12 for Twitter and 40 for Weibo), rendering a precision of 94.8%. Among the left 948 true SMS spam screenshots, our STR has accurately recognized the spam text for 895 screenshots, which renders an accuracy of 94.4%. We then evaluated the performance in terms of *word accuracy* and *character accuracy*, where the former is the ratio of correctly recognized SMS words to all identified words, and the latter is the ratio of correctly identified characters to all detected characters. STR achieves a word accuracy of 97% and a character accuracy of 99%.

# Appendix B.
# More Details on Understanding SMS Spam Messages

TABLE 16. A PERFORMANCE COMPARISON OF BINARY SMS SPAM CLASSIFICATION MODELS.

| Model | Test$_{All-Twitter}$ | | Test$_{All-Weibo}$ | |
|---|---|---|---|---|
| | Precision | Recall | Precision | Recall |
| CNN$_{All}$ | 97.92% | 98.33% | 97.90% | 98.54% |
| BERT$_{All}$ | **98.72%** | 99.40% | **98.48%** | 99.29% |
| BERT$_{All-Twitter}$ | 97.67% | **99.59%** | 97.23% | 98.82% |
| BERT$_{All-Weibo}$ | 98.64% | 99.25% | 98.40% | **99.40%** |

TABLE 17. THE PERFORMANCE OF MULTI-LABEL SMS SPAM CLASSIFICATION.

| Spam Category | % Groundtruth | Precision | Recall | F1-Score |
|---|---|---|---|---|
| Promotion | 49.93% | 85.71% | 90.45% | 87.72% |
| P-Gambling | 1.49% | 40.00% | 80.00% | 53.33% |
| P-Sex | 1.64% | 85.71% | 100.00% | 92.31% |
| F-Finance | 33.91% | 87.92% | 85.62% | 86.75% |
| F-Account | 28.26% | 80.17% | 82.20% | 81.17% |
| F-Insurance | 1.83% | 100.00% | 100.00% | 100.00% |
| F-Delivery | 7.24% | 92.86% | 86.67% | 89.66 % |
| F-Acquaintance | 1.09% | 100.00% | 33.33% | 50.00% |
| F-COVID | 1.44% | 100.00% | 66.67% | 80.00% |
| F-Prize | 5.45% | 81.82% | 47.37% | 60.00% |

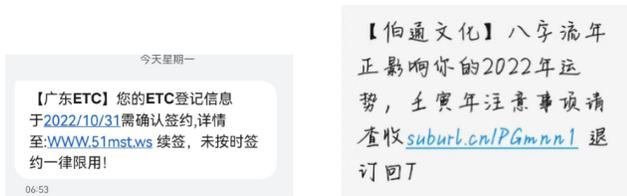

(a) A SMS spam masquerading as a service update message of electronic toll collection (ETC).

(b) A SMS spam promoting a fortune-telling service.

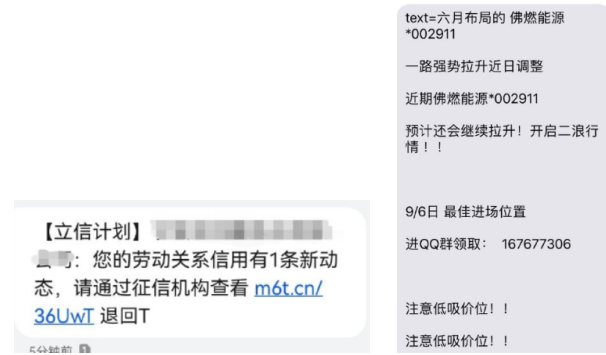

(c) A SMS spam masquerading as an update message from a credit bureau.

(d) A SMS spam involving securities fraud and stock manipulation.

Figure 7. Interesting SMS spam cases as reported on the Weibo platform.

Also, many Weibo-reported spam messages are composed with text semantics that are closely related to Chinese culture or local services in China. Several interesting cases are presented in Figure 7. For example, Figure 7(a) presents a spam message about ETC, an electronic toll collection system that is gradually becoming popular in China, while the case in Figure 7(b) is to promote a Chinese fortune-telling service, which can easily attract some people who believe in divination. Besides, novel fraud activities have also been observed, e.g., phishing attacks relevant to financial credit (Figure 7(c)), and the promotion of illegal stock manipulation(Figure 7(d)).

# Appendix C.
# More Details on SMS Spam Classification

## C.1. Performance Results of the Binary SMS Spam Classification

Table 16 lists the performance of SMS spam binary classification models when evaluated on Test$_{All-Twitter}$ and Test$_{All-Weibo}$.

## C.2. The Prompt of GPT-4

From now on, you are a researcher in cybercrime and Spam detection task. You will be provided with a text and then follow the steps below to perform the tasks. In the end, you just need to return a JSON data that fulfills the requirements.

Step 1: Determine whether the text is relevant to sms(short message service) spam text or not.

Step 2: Following the result of Step 1, if the text is not relevant to sms spam at all, the result of this step is 'benign'. Otherwise, the result of this step is 'spam'.

Step 4: Gather the results before and generate the output JSON. The JSON data has 1 required fields, 'Spam'. The 'Spam' field is the result from Step 2, and it must be 1 or 0, with the 0 representing Benign and 1 representing spam.

Here is the text: "'TEXT"'.

## C.3. The Performance of Multi-Label SMS Spam Classifier

We also measured the category-wise precision and recall for the multi-label SMS spam classifier, as shown in Table 17.

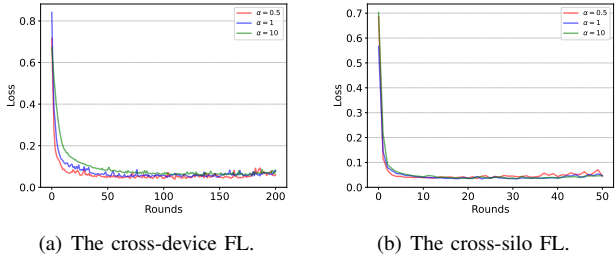

(a) The cross-device FL.   (b) The cross-silo FL.

Figure 8. The convergence process of FL-trained SMS spam detection models.

TABLE 18. THE ROBUSTNESS OF THE ADVERSARIALLY TRAINED SMS SPAM DETECTION MODEL WHEN EVALUATED AGAINST THREE IMPERCEPTIBLE ATTACKS.

| Model / Attack Methods | Homoglyph | Reorder | Deletion |
| --- | --- | --- | --- |
| BERT$_{All}$ | 0.084 | 0.168 | 0.200 |
| BERT$_{adversarial,All}$ | 0.000 | 0.044 | 0.048 |

## C.4. The FL Training Process of SMS Spam Detection Models

Figure C.3 presents the training process of these FL models, and it takes up to 59/82/127 rounds to converge for cross-device FL scenarios when $\alpha \in \{0.5, 1, 10\}$ respectively, while it is only 32/33/41 rounds for cross-silo FL when $\alpha \in \{0.5, 1, 10\}$.

# Appendix D.
# More Results on the Adversarial Resistance of SMS Spam Classification

## D.1. More Details on the Effectiveness of Adversarial Training

Table 18 presents a direct comparison between the original SMS spam detection model (BERT$_{All}$) and the adversarially trained counterpart (BERT$_{adversarial,All}$) with regards to their adversarial resistance against three imperceptible attacks of adversarial examples.

## D.2. More Results on the Untargeted Poisoning Attacks

Figure 9 illustrate how the spam detection model varies in its performance under untargeted poisoning attacks of different poisoning rates.

## D.3. More Details on the Reverse Backdoor Attacks

Regarding the reverse backdoor attack via reporting stamped spam messages (the poisoning scenario III), Figure 10(a) presents the backdoor effect in terms of misclassifying benign testing messages that are stamped with

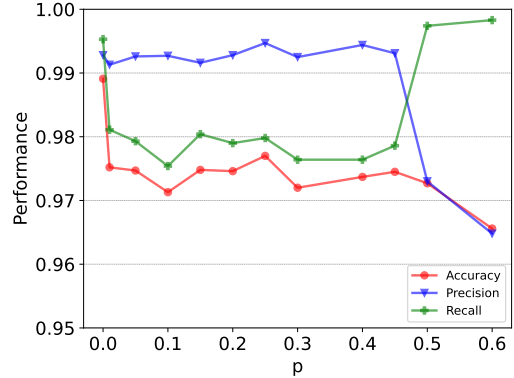

Figure 9. The attacking impact for untargeted poisoning attacks under different poisoning rates $p$.

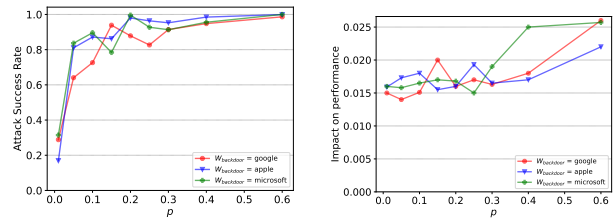

(a) The attack success rate.   (b) The impact on the model performance (the decrease in accuracy).

Figure 10. The effectiveness of reverse backdoor attacks via reporting stamped true **spam** messages.

the respective backdoor word, while Figure 10(b) presents the impact of these backdoor attacks on the overall model performance.

# Appendix E.
# The Effect of Concept Drift

Given adversarial robustness evaluated for SSD models, we step further to profile another robustness issue, namely, concept drift. Concept drift refers to the evolution of data distribution and the resulting effect in terms of changing the statistical data properties, which can in turn invalidate a machine learning model built upon the data. When concept drift occurs, a machine learning classifier can have a decaying performance when predicting on emerging and future data samples. Previous studies [65]–[67] have profiled the issue in many classification tasks, e.g., computer vision [65], [66] and NLP [67]. This is even the case for tasks in the security domain [12], [68] since the miscreants have strong incentives to proactively change the data distribution so as to escape from detection. In this part, we proceed to evaluate the effect of concept drift on SMS spam detection. Previously, this is not doable since none of publicly available spam datasets contain timing metadata for the included SMS spam messages. However, since our spam dataset is collected from spam-reporting OSN posts and each has the reporting time attached. Assuming each of these SMS spam messages was reported shortly after the victim had received it, the temporal distribution of spam messages across their reporting time can serve as a good approximation for their authentic temporal distribution. Such a temporal distribution gives us the opportunity, to evaluate for the first time, the impact of concept drift on SSD models, i.e., how a spam classification model decays across time in its performance when predicting future spam messages.

**The experiment setting.** In a nutshell, our concept drift experiments involve two steps. One is to train a set of binary SMS spam models while the second is to evaluate them against future spam messages. Specifically, a set of binary SMS spam models {$Model_{year}$} have been trained, which differ in the time range of the training data. For instance, $Model_{2018}$ refers to the model trained on spam samples observed before the end of 2018, while $Model_{2015}$ is trained on spam samples no later than 2015. Among the four datasets listed in Table 4.1, the UCI was collected in 2012 and the ExAIS was collected in 2015, while both Twitter and Weibo datasets cover the time period between January, 2018 and March 2023. We thus trained four models, namely, $Model_{2012}$, $Model_{2015}$, $Model_{2018}$, $Model_{2021}$. Among these four models, $Model_{2012}$ is trained solely on 80% of the UCI dataset while the left are used for testing, $Model_{2015}$ is built upon both the UCI and ExAIS with 80% for training and 20% for testing, while $Model_{2018}$ extends the dataset to spam messages observed before 2019 on both Weibo and Twitter and $Model_{2021}$ further includes spam samples reported between 2019 and 2021. Also, all these four models share a model architecture and training parameters the same as the binary multilingual classification model ($BERT_{All}$) introduced in §4. To evaluate the effect of concept drift on these models, all spam messages observed on both Twitter and Weibo are divided into groups by the quarter of the year when each of them was observed, e.g., a spam message reported on February 2019 will be assigned to the group of *2019 Q1*. Then, each model will be evaluated on both these quarter-based spam groups as well as the testing segments of the UCI dataset and the ExAIS dataset.

Table 19 presents the performance of these four models in terms of recall, when evaluated against spam datasets of different time periods. As you can see, the decay in performance varies significantly across models. Particularly, $Model_{2012}$ and $Model_{2015}$ have decayed significantly in the recall performance when predicting spam messages of 2018 or later. For instance, $Model_{2012}$ has achieved a initial recall of 94.85% when evaluated on the testing part of the UCI dataset. However, its performance degrades to 77.67%, 82.6%, and 70.97% when evaluated respectively on spams observed in 2019Q1, 2021Q1, and 2022 Q1. On the other hand, the two models trained on less outdated datasets are subject to less degradation in performance. Particularly, compared to $Model_{2012}$, $Model_{2018}$ has achieved a recall higher by 27.72% when evaluated on 2022Q1 and the increase in recall is 28.88% for $Model_{2021}$. These results not only qualify the existence of concept drift but also quantify its impact in model aging in the area of SMS spam detection, which strongly highlights the necessity of maintaining an ever-updating SMS spam dataset and keeping retraining the detection models on fresh spam messages.

TABLE 19. THE CONCEPT DRIFT OF SMS SPAM DETECTION MODELS IN TERMS OF RECALL WHEN EVALUATED ON FUTURE SMS SPAM DATASETS.

| Model | UCI | ExAIS | 2018Q1 | 2019Q1 | 2020Q1 | 2021Q1 | 2022Q1 | 2023Q1 |
|---|---|---|---|---|---|---|---|---|
| $Model_{2012}$ | 99.03% | 62.56% | 76.12% | 77.67% | 60.34% | 82.60% | 70.97% | 78.13% |
| $Model_{2015}$ | 97.87% | 88.71% | 72.64% | 68.41% | 52.29% | 73.48% | 66.18% | 70.27% |
| $Model_{2018}$ | 98.07% | 89.47% | 99.00% | 99.05% | 98.77% | 98.65% | 98.69% | 98.28% |
| $Model_{2021}$ | 97.10% | 88.20% | 99.50% | 100% | 99.66% | 99.49% | 99.85% | 100% |

# Appendix F.
# The Transferability of SMS Spam Detection Models

**SMS/Email spam classification.** To characterize the transferability between SMS spam classification and Email spam classification, a set of binary spam classification model have been built up and evaluated against different testing datasets. As listed in Table 4.1, these models share a model architecture and training parameters the same as the original binary SMS spam classifier (§4.1), but differ in their respective training dataset and training process. Regarding the training dataset, some models (e.g., $Model_{SMS}$ and $Model_{Email}$) are trained on either the SMS spam training dataset $Train_{SMS}$ or the Email spam training dataset $Train_{Email}$, while the other models are trained on both datasets which include $Model_{SMS\&Email}$, $Model_{SMS \to Email}$, and $Model_{Email \to SMS}$. Concerning the training process, most models follow the regular training process except for $Model_{SMS \to Email}$, and $Model_{Email \to SMS}$. For $Model_{SMS \to Email}$, it is first trained on $Train_{SMS}$ before being fine-tuned on $Train_{Email}$ so as to evaluate whether knowledge distilled from SMS spam data can boost the model performance for Email spam classification. On the other hand, $Model_{Email \to SMS}$ shares a similar training process but reverses the order of the training datasets.

Then, regarding the datasets, $Train_{SMS}$ refers 80% of all the SMS spam messages we have collected while the left are used as the testing dataset $Test_{SMS}$. Besides, for Email spam, $EnronSpam$, a public Email dataset [12], is utilized to compose $Train_{Email}$ and $Test_{Email}$ following the same ratio (80%:20%). To give you a fresh impression regarding the data size, $EnronSpam$ contains 15581 non-spam Emails and 14566 spam Emails.

Table 20 presents the performance of these models when evaluated against three different testing datasets, namely, $Test_{SMS}$, $Test_{Email}$, and the combination of both. As we can see, the model trained solely on SMS spam data ($Model_{SMS}$) achieves a recall of 85.00% and a precision of 55.03% when evaluated on $Test_{Email}$, which is likely due to that SMS spam and Email spam still differ in their respective data distribution. However, When evaluated on

12. http://nlp.cs.aueb.gr/software_and_datasets/Enron-Spam/index.html

TABLE 20. A DIRECT COMPARISON OF VARIOUS SPAM DETECTION MODELS.

| Model | Test$_{SMS}$ Prec. | Test$_{SMS}$ Recall | Test$_{Email}$ Prec. | Test$_{Email}$ Recall | Test$_{SMS}$ & Test$_{Email}$ Prec. | Test$_{SMS}$ & Test$_{Email}$ Recall |
|---|---|---|---|---|---|---|
| Model$_{SMS}$ | 99.28% | 99.53% | 55.13% | 85.03% | 87.92% | 96.61% |
| Model$_{Email}$ | 91.57% | 78.62% | 99.68% | 94.30% | 92.58% | 82.14% |
| Model$_{SMS\&Email}$ | 99.15% | 99.48% | 98.69% | 99.37% | 99.10% | 99.40% |
| Model$_{SMS \rightarrow Email}$ | 97.38% | 97.02% | 99.03% | 99.10% | 97.78% | 97.41% |
| Model$_{Email \rightarrow SMS}$ | 99.21% | 99.18% | 76.80% | 91.35% | 94.62% | 97.57% |

Test$_{Email}$, Model$_{SMS\&Email}$ achieves a comparable precision and a higher recall of 99.35% while it is only 94.34% for Model$_{Email}$. The same performance improvement is also observed for Model$_{SMS \rightarrow Email}$, which suggests the SMS spam dataset can benefit the detection of Email spam while the reverse effect from Email spam to SMS spam is not obvious.

**SMS spam classification and general toxic text classification.** The toxic content dataset used in our experiments is IBM toxic text dataset [13] wherein a toxic text will be assigned with one or more toxic labels. In our experiments, a text in this dataset will be considered as toxic (positive) if it has been assigned one or more toxic labels. As the result, we got a binary dataset of 201,081 non-toxic texts and 22,468 toxic texts. Adopting the same model architecture and training process of BERT$_{All}$ (the binary SMS spam classifier), a model trained on 80% of this toxic text dataset has achieved a precision of 61.09%, a recall of 79.43% and a f1-score of 69.06%, when evaluated on the left 20% texts (the toxic testing dataset). We then tried to apply the SMS spam detection model BERT$_{All}$ to predicting the toxic testing dataset. However, it only achieved a precision of 5.55%, a recall of 30.21% and a F1-score of 9.38%, which means the poor portability of SMS spam classifier on toxic text classification. Reversely and similarly, when evaluated on the SMS spam testing dataset, the toxic binary classifier also achieved a poor performance (a precision of 53.03%, a recall of 0.47% and the F1-score of 0.94%).

We then carefully analyzed the content of toxic text and SMS spam, and found significant difference in terms of text semantics. Particularly, the main factor for determining whether or not a text is toxic is to check the existence of foul language, discriminatory words, and negative tones, while the key factor for determining SMS spam is whether the content is related to fraud, promotion, and so on.

---

13. https://www.ml-exchange.org/models/max-toxic-comment-classifier/